\documentclass[12pt]{article}
\usepackage{jheppub}
\pdfoutput=1

\usepackage{amsmath,bbm,array,amsfonts,graphicx,wrapfig,lscape,float,mathtools,multirow,longtable}
\usepackage{youngtab}

\newcommand{\be}{\begin{equation}}
\newcommand{\ee}{\end{equation}}
\newcommand{\beq}{\begin{equation}}
\newcommand{\beql}[1]{\begin{equation}\label{#1}}
\newcommand{\eeq}{\end{equation}}
\newcommand{\ba}{\begin{array}}
\newcommand{\ea}{\end{array}}
\newcommand{\beal}[1]{\begin{eqnarray}\label{#1}}
\newcommand{\bea}{\begin{equation} \begin{aligned}} \newcommand{\eea}{\end{aligned} \end{equation}}
\newcommand{\ben}{\begin{enumerate}}
\newcommand{\een}{\end{enumerate}}
\newcommand{\bean}{\begin{eqnarray*}}
\newcommand{\eean}{\end{eqnarray*}}

\newcommand{\nn}{\nonumber}

\newcommand{\fref}[1]{Figure \ref{#1}}
\newcommand{\btab}[1]{\begin{tabular}{#1}}
\newcommand{\etab}{\end{tabular}}

\def \Black {\pdfsetcolor{0 0 0 1}}

\newcommand{\comment}[1]{}

\newcommand{\CM}{{\cal M}}

\newcommand{\CN}{{\cal N}}

\newcommand{\qed}{\nobreak \ifvmode \relax \else
      \ifdim\lastskip<1.5em \hskip-\lastskip
      \hskip1.5em plus0em minus0.5em \fi \nobreak
      \vrule height0.75em width0.5em depth0.25em\fi}

\newcommand{\Tr}{\text{Tr}}

\newcommand{\ov}{\over}
\newcommand{\GG}{\mathbf{G}}

\newcommand{\m}{\mathfrak{m}}
\newcommand{\n}{\mathfrak{n}}
\newcommand{\Z}{\mathbb{Z}}
\newcommand{\C}{\mathbb{C}}
\newcommand{\R}{\mathbb{R}}
\renewcommand{\t}{\widetilde }
\renewcommand{\d}{\partial }

\newcommand{\CA}{\mathcal{A}}

\newcommand{\CO}{\mathcal{O}}
\newcommand{\CQ}{\mathcal{Q}}
\newcommand{\CV}{\mathcal{V}}
\renewcommand{\b}{\bar }

\newcommand{\FR}{\mathfrak{R}}
\newcommand{\Fg}{\mathfrak{g}}

\newcommand{\half}{{1\over 2}}

\title{Supersymmetric gauged matrix models from dimensional reduction on a sphere}

\author[a]{Cyril Closset,}
\author[b]{Dongwook Ghim,}
\author[c]{Rak-Kyeong Seong}

\affiliation[a]{
Theory Department, CERN, CH-1211, Geneva 23, Switzerland
}

\affiliation[b]{
Department of Physics and Astronomy, Seoul National University, Seoul 08826, Korea
}

\affiliation[c]{
Yau Mathematical Sciences Center, Tsinghua University, 100084 Beijing, China
}

\emailAdd{cyril.closset@cern.ch}
\emailAdd{sg1841@snu.ac.kr}
\emailAdd{rakkyeongseong@gmail.com}

\preprint{
\begin{flushright}
CERN-TH-2017-284\\
SNUTP17-005\\
\end{flushright}
}

\abstract{It was recently proposed that $\mathcal{N}=1$ supersymmetric gauged matrix models have a duality of order four---that is, a quadrality---reminiscent of infrared dualities of SQCD theories in higher dimensions. In this note,  we show that the zero-dimensional quadrality proposal can be infered from the two-dimensional Gadde-Gukov-Putrov triality.  We consider two-dimensional $\CN=(0,2)$ SQCD compactified on a sphere with the half-topological twist. For a convenient choice of $R$-charge, the zero-mode sector on the sphere gives rise to a simple $\mathcal{N}=1$ gauged matrix model. Triality on the sphere then implies a triality relation for the supersymmetric matrix model, which can be completed to the full quadrality.
}

\begin{document}

\maketitle

\section{Introduction}

Supersymmetric quantum field theories in various space-time dimensions can be related to each other in a number of way, which often leads to fruitful new perspectives on their dynamics. In this note, we are interested in a highly degenerate case: a ``field theory'' in zero dimension---that is, a matrix model. More precisely, we are interested in gauged matrix models with one supersymmetry, as recently studied by Franco, Lee, Seong and Vafa  \cite{Franco:2016qxh, Franco:2016tcm}. Such models arise naturally from $D$-instantons in type IIB string theory; in particular,  D$(-1)$-branes at the tip of a Calabi-Yau fivefold singularity  preserve $0d$ $\CN=1$ supersymmetry, and it is expected that their low-energy dynamics is captured by a quiver $\CN=1$ gauged matrix model---see in particular \cite{Franco:2016tcm, Franco:2017lpa, Closset:2017yte} for recent studies.

The simplest gauged matrix model is a zero-dimensional version of SQCD. It consists of a $U(N_c)$ gauge group~\footnote{A $0d$ gauge symmetry is an invariance of the matrix model. While there are no $0d$ gauge fields, the $0d$ $\CN=1$ vector multiplet contains a gaugino and an auxiliary field $D$, which leads to familiar $D$-term constraints \cite{Franco:2016tcm}.} with bosonic and fermionic matrices transforming in the fundamental or antifundamental representations of $U(N_c)$. The model has a flavor symmetry group:
\be
\GG_F={U(N_1)\times U(N_2)\times U(N_3)\times U(N_4) \ov U(1)}~,
\ee
such that $N_1-N_2+N_3-N_4=0$.
It was  proposed in \cite{Franco:2016tcm} that $0d$ SQCD has four equivalent descriptions, as we will review momentarily. This {\it quadrality} permutes the four flavor groups $U(N_i)$ in $\GG_F$.

The quadrality proposal is on somewhat weaker footing than analogous dualities in quantum field theory in dimensions $d\geq 2$.  Zero-dimensional SQCD is defined as a formal supersymmetric integral. The integral itself (the ``partition function'') generally vanishes due to ```t Hooft anomalies,'' and the non-vanishing observables have not  yet been computed explicitly. Nonetheless, we will argue that the quadrality proposal arises naturally as a consequence of the Gadde-Gukov-Putrov triality for two-dimensional $\CN=(0,2)$ SQCD \cite{Gadde:2013lxa}. 

Supersymmetric field theories in various dimensions can often be related by dimensional reduction: one obtains a $d$-dimensional theory from a $(d+n)$-dimensional theory by supersymmetric compactification on some $n$-manifold $\CM_n$, by sending the size $R$ of $\CM_n$ to zero.
While one can always perform such reductions at the level of the classical Lagrangian, renormalization group flows across dimensions can be rather subtle  \cite{Niarchos:2012ah, Aharony:2013dha, Aharony:2013kma, Benini:2015bwz, Hwang:2017nop, Aharony:2017adm}. For instance, in the case of the $S^1$ reduction of theories with four supercharges, the $R\rightarrow 0$ limit does not commute with the infrared (IR) limit \cite{Aharony:2013dha, Aharony:2017adm}. There is no guarantee, in general, that infared dualities between $(d+n)$-dimensional theories lead to infared dualities between their dimensionally-reduced $d$-dimensional cousins.

To avoid this issue, one may consider a compactification with a topological twist. By construction,  observables of the (partially) topologically-twisted theory should be independent of the size $R$ of the compactification manifold, and therefore one should be able to safely flow to the IR. A prime example of this procedure is the compactification of $4d$ $\CN=1$ gauge theories on $\R^2 \times S^2$ with a topological twist on $S^2$ \cite{Closset:2013sxa, Benini:2015noa, Honda:2015yha, Gadde:2015wta}. This preserves $2d$ $\CN=(0,2)$ supersymmetry along $\R^2$. Using this setup, it has been argued that $\CN=1$ Seiberg duality in $4d$ implies $\CN=(0,2)$ triality in $2d$ \cite{YujiUnpublished, Honda:2015yha, Gadde:2015wta}. 

To obtain $0d$ SQCD, we similarly consider $2d$ $\CN=(0,2)$ SQCD on $S^2$ with a half-topological twist. This compactification preserves the right-moving $R$-symmetry $U(1)_R$ of the 2d theory, by turning on a non-trivial background $U(1)_R$ gauge field:
\be
{1\ov 2 \pi} \int_{S^2} d A^{(R)}= -1~.
\ee
By a convenient choice of $R$-charge, and restricting to the zero-modes---the lowest modes on $S^2$ with one unit of $U(1)_R$ flux---we directly reproduce the $0d$ SQCD studied in \cite{Franco:2016tcm}. More precisely, this holds in a topological sector with vanishing gauge flux. (We only briefly comment on the dimensionally-reduced theory in the non-trivial topological sectors, where the gauge group is broken explicitly to a Levi subgroup. In the reduction of \cite{Gadde:2015wta}, they could argue that only the zero-flux sector contributed to some $S^2\times T^2$ partition function. We expect something similar to happen here, for certain observables.) Three of the four dual formulations of $0d$ SQCD follow from the conjectured triality in $2d$. The fourth formulation of the gauged matrix model can be recovered by consistency.

\vskip0.3cm
This note is organized as follows. In section  \ref{2d review}, we review how triality maps certain $2d$ $\CN=(0,2)$ supersymmetric gauge theories. In section \ref{matrix models}, we discuss gauged matrix models with $\CN=1$ supersymmetry, and we review the quadrality proposal; next, we  review how the zero-modes of $2d$ $\CN=(0,2)$ multiplets on the half-twisted sphere sit in $0d$ supersymmetric multiplets; we then show that,  for a convenient choice on $R$-charge, $2d$ SQCD can be reduced to $0d$ SQCD, and $2d$ triality descends to $0d$ quadrality. We also briefly discuss a related setup with D-branes at Calabi-Yau singularities.
Our conventions are summarized in Appendix.



\section{Two-dimensional $\CN=(0,2)$ SQCD and triality}\label{2d review}

Consider a two-dimensional $\CN=(0,2)$ gauge theory with $U(N_c)$ gauge group, ${\bf N}_1$ chiral multiplets $\Phi_{{\tiny\yng(1)}}$ in the fundamental representation, ${\bf N}_2$ chiral multiplets $\Phi_{\overline{\tiny\yng(1)}}$ in the anti-fundamental representation, ${\bf N}_3$ fermi multiplets $\Lambda_{{\tiny\yng(1)}}$   in the fundamental representation, and two fermi multiplets $\Omega_\pm$ in the ${\rm {\bf det}}^{\pm1}$ representation of $U(N_c)$, as summarized in Table \ref{tab: SQCD fields}. All the fermi multiplets have vanishing $E$- and $J$-term superpotentials. We must have:
\be
N_c= {{\bf N}_1+ {\bf N}_2- {\bf N}_3\ov 2}~,
\ee
in order to cancel the gauge anomalies.

\begin{figure}[t]
\begin{center}
\resizebox{0.9\hsize}{!}{
\includegraphics[width=8cm]{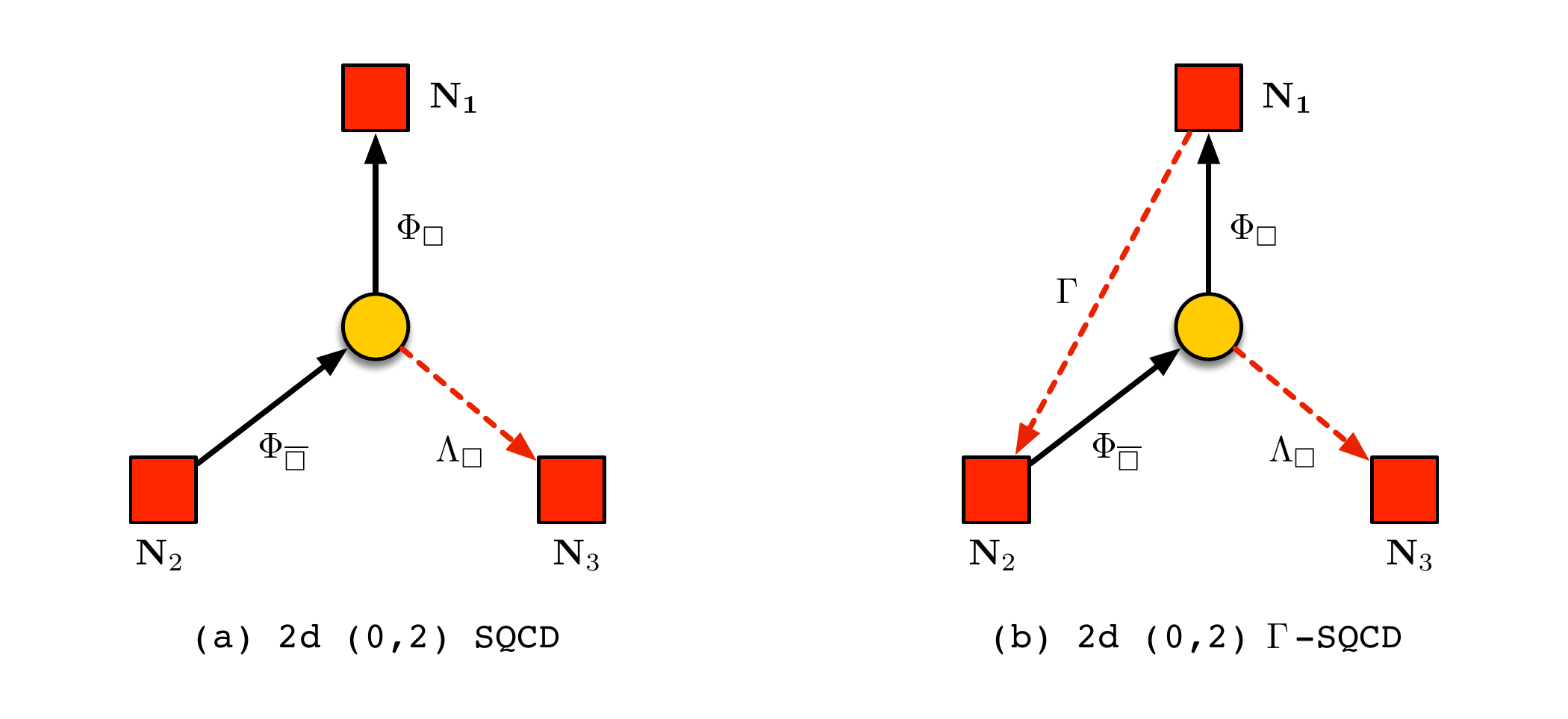}
}
\caption{
The quiver diagram of $2d$ $\CN=(0,2)$ (a) SQCD and (b) $\Gamma-$SQCD.  The fermi fields $\Omega_\pm$ in the determinant representation of $U(N_c)$ are omitted to avoid clutter.
\label{2d02sqcd}}
 \end{center}
 \end{figure}

The flavor group of the theory is:
\be\label{GGF 2d}
\GG_F= {U({\bf N}_1) \times U({\bf N}_2) \times U({\bf N}_3)\ov U(1)}~.
\ee
Here the quotient is by the overall $U(1) \subset U(N_c)$ gauge symmetry.
The theory also has a right-moving $R$-symmetry $U(1)_R$, which assigns $R$-charge $\pm 1$ to the supercharges $\b \CQ_+$ and $\CQ_+$, respectively. For future reference, it is useful to allow generic $R$-charges, as indicated in Table \ref{tab: SQCD fields}, which generally breaks the flavor group \eqref{GGF 2d} to its maximal torus.
The vanishing of the mixed $U(1)_R$-gauge anomaly requires:
\be
\sum_{i=1}^{{\bf N}_1} (r_i-1)- \sum_{j=1}^{{\bf N}_2} (\t r_j-1)- \sum_{I=1}^{{\bf N}_3}  r_I - N_c r_+ +N_c r_{-} =0~.
\ee
The flavor indices $i, j, I$ run over the $U({\bf N}_1)$, $U({\bf N}_2)$ and $U({\bf N}_3)$ flavor groups, respectively.
\begin{table}[t]
\centering
\be\nn
\begin{array}{c|c|cccc}
    & U(N_c)& U({\bf N}_1) & U({\bf N}_2) & U({\bf N}_3)   & U(1)_R  \\
\hline
{\Phi_{{\tiny\yng(1)}}}_{\,i} & {\bf N_c} &  {\bf \overline N_1} & {\bf 1}&{\bf 1}& r_i\\
{\Phi_{\overline{\tiny\yng(1)}}}^{\,j}  & {\bf \overline N_c} &  {\bf 1} & {\bf N_2}&{\bf 1}& {\t r_j}\\
{\Lambda_{{\tiny\yng(1)}}}_{\,I} & {\bf N_c} &  {\bf 1} & {\bf 1}&{\bf \overline N_3}& r_I \\
\Omega_\pm & {\bf det}^{\pm 1} &  {\bf 1} & {\bf 1}&{\bf 1}& r_\pm  \\
\hline
{\Gamma^i}_j & {\bf 1} & {\bf N_1}& {\bf \overline N_2}& {\bf 1}& r_\Gamma
\end{array}
\ee
\caption{Charges of the matter fields in $2d$ $\CN=(0,2)$ SQCD. The ${\bf det}$ and $ {\bf det}^{-1}$ representation assigns charge $+1$ and $-1$, respectively, to each $U(1)_a \subset U(N_c)$, $a=1, \cdots, N_c$, in the Cartan subgroup of $U(N_c)$. The last line is the fermi gauge-singlet in $\Gamma$-SQCD}
\label{tab: SQCD fields}
\end{table}
Note that our definition of $2d$ SQCD differs slightly from the one \cite{Gadde:2013lxa}. The original definition also involves ${\bf N}_1 {\bf N}_2$ additional  gauge-singlet fermi multiplets ${\Gamma^i}_j$ with:
\be\label{superpot GammaSQCD}
E_{{\Gamma^i}_j}= 0~, \qquad \qquad 
J_{{\Gamma^i}_j}=  {\Phi_{\overline{\tiny\yng(1)}}}^{\, j}\, {\Phi_{{\tiny\yng(1)}}}_{\, i}~, 
\ee
for the $\CN=(0,2)$ superpotential, where the gauge indices are kept implicit. This superpotential ensures that, in the appropriate range of parameters, the theory flows to an SCFT with a normalizable vacuum. We will call this theory, with the gauge-singlet fermi multiplets added, ``$\Gamma$-SQCD''. The superconformal $R$-charges of $\Gamma$-SQCD were determined in \cite{Gadde:2013lxa} using $c$-extremization \cite{Benini:2012cz}. Both types of $\mathcal{N}=(0,2)$ SQCD's we discussed so far are summarized in \fref{2d02sqcd}.

\subsection*{First triality move:}
It was argued in \cite{Gadde:2013lxa} that 2d $\CN=(0,2)$  SQCD has two other dual descriptions, forming a so-called {\it triality} of gauge theories with identical infrared physics. The first dual theory is a $U(N_c')$ theory with:
\be
N_c'= {\bf N}_2-N_c =  {{\bf N}_2+ {\bf N}_3- {\bf N}_1\ov 2}~,
\ee
and the matter content:
\be\label{2d SQCD dual1}
\begin{array}{c|c|cccc}
    & U(N_c')& U({\bf N}_1) & U({\bf N}_2) & U({\bf N}_3)   & U(1)_R  \\
\hline
{\Lambda'_{{\tiny\yng(1)}}}_{\,i} & {\bf N_c'} &  {\bf \overline N_1} & {\bf 1}&{\bf 1}& r'_i\\
{\Phi'_{{\tiny\yng(1)}}}_{\,j}  & {\bf  N_c'} &  {\bf 1} & {\bf \overline N_2}&{\bf 1}& {\t r'_j}\\
{\Phi'_{\overline{\tiny\yng(1)}}}^{\,I} & {\bf \overline {N_c'}} &  {\bf 1} & {\bf 1}&{\bf N_3}& r'_I \\
\Omega'_\pm & {\bf det}^{\pm 1} &  {\bf 1} & {\bf 1}&{\bf 1}& r'_\pm  \\
{M^j}_i & {\bf 1} & {\bf  \overline N_1}& {\bf N_2}& {\bf 1}& r_M\\
{\Gamma'^{\,j}}_I & {\bf 1}& {\bf 1}& {\bf N_2}& {\bf \overline N_3}& r'_\Gamma
\end{array}
\ee
The duality operation changes the gauge group, and permutes the (anti)fundamental matter according to:
\be\label{trialityrule}
{\Phi_{\overline{\tiny\yng(1)}}} \;\longrightarrow\; {\Phi'_{{\tiny\yng(1)}}}~, \qquad 
{\Phi_{{\tiny\yng(1)}}} \;\longrightarrow\; {\Lambda'_{{\tiny\yng(1)}}}~, \qquad 
{\Lambda_{{\tiny\yng(1)}}} \;\longrightarrow\; {\Phi'_{\overline{\tiny\yng(1)}}}~.
\ee
The dual theory also contains ``mesonic''  gauge-singlet, the chiral multiplets ${M^j}_i$ and the fermi multiplets ${\Gamma'^{\,j}}_I$, which are identified with gauge invariant operators in the original theory:
\be
{M^j}_i = {\Phi_{\overline{\tiny\yng(1)}}}^{\,j}\, {\Phi_{{\tiny\yng(1)}}}_{\,i}~, \qquad\quad
{\Gamma'^{\,j}}_I =  {\Phi_{\overline{\tiny\yng(1)}}}^{\,j}\, {\Lambda_{{\tiny\yng(1)}}}_{\,I}~.
\ee
Importantly, the fermi multiplets of the dual theory have non-trivial superpotentials:
\bea
&  E_{\Lambda'_i} ={\Phi'_{{\tiny\yng(1)}}}_{\,j}  \, {M^j}_i~, \qquad \qquad && J_{\Lambda'_{i}}=0~, \cr
& E_{{\Gamma'^{\,j}}_I }=0~, \qquad \qquad && J_{{\Gamma'^{\,j}}_I } =  {\Phi'_{\overline{\tiny\yng(1)}}}^{\,I}\, {\Phi'_{{\tiny\yng(1)}}}_{\,j} ~.
\eea 
The duality operation can be conveniently summarized in terms of quiver diagrams, as shown in Figure \ref{trialitycycle}.

It is useful to think of this duality operation as a ``local'' operation on the gauge group $U(N_c)$, which might be part of a more general theory. In particular, there might be additional superpotential terms of $E_\Lambda$ and $J_\Lambda$ in the original theory. In that case, the duality move introduces the interaction terms \cite{Franco:2017lpa, Closset:2017yte}:
\bea\label{duality operation}
&  E_{\Lambda'_i} ={\Phi'_{{\tiny\yng(1)}}}_{\,j}  \, {M^j}_i~, \qquad \qquad && J_{\Lambda'_{i}}=\left({\d E_{\Lambda_I} \ov \d { {\Phi_{{\tiny\yng(1)}}}_{\, i}}} \right){\Phi'_{\overline{\tiny\yng(1)}}}^{\,I}~, \cr
& E_{{\Gamma'^{\,j}}_I }=- {M^j}_i \left({\d   E_{\Lambda_I}\ov \d  {{\Phi_{{\tiny\yng(1)}}}_{\, i}}} \right)~, \qquad \qquad && J_{{\Gamma'^{\,j}}_I } =  {\Phi'_{\overline{\tiny\yng(1)}}}^{\,I}\, {\Phi'_{{\tiny\yng(1)}}}_{\,j} \; -\, {\d   J_{\Lambda_I}\ov \d {\Phi_{\overline{\tiny\yng(1)}}}^{\,j} }~,
\eea 
where we sum over repeated indices. The determinant fields $\Omega_\pm$ are spectators in these dualities.

The dual theory for $\Gamma$-SQCD is similar, albeit simpler. After this duality move, the gauge-singlet fields $\Gamma$ and $M$ are massive, due to the superpotential \eqref{superpot GammaSQCD}, which becomes $J_\Gamma =M$ after the duality. After integrating them out, the dual theory looks identical to the original $\Gamma$-SQCD theory, up to the reshuffling of the flavor parameters $({\bf N}_1, {\bf N}_2, {\bf N}_3)\rightarrow ({\bf N}_2, {\bf N}_3, {\bf N}_1)$.

\subsection*{Second triality move:}
By applying the duality operation \eqref{trialityrule}-\eqref{duality operation} a second time, to the theory \eqref{2d SQCD dual1}, we find another ``dual'' theory. This is a $U(N_c'')$ theory with:
\be
N_c''= {\bf N}_3-N_c' =  {{\bf N}_3+ {\bf N}_1- {\bf N}_2\ov 2}~,
\ee
and the following matter content:
\be\label{2d SQCD dual2}
\begin{array}{c|c|cccc}
    & U(N_c'')& U({\bf N}_1) & U({\bf N}_2) & U({\bf N}_3)   & U(1)_R  \\
\hline
{\Phi''_{\overline{\tiny\yng(1)}}}^{\,i} & {\bf \overline{N_c''}} &  {\bf N_1} & {\bf 1}&{\bf 1}& r''_i\\
{\Lambda''_{{\tiny\yng(1)}}}_{\,j}  & {\bf  N_c''} &  {\bf 1} & {\bf \overline N_2}&{\bf 1}& {\t r''_j}\\
{\Phi''_{{\tiny\yng(1)}}}_{\,I} & {\bf N_c''} &  {\bf 1} & {\bf 1}&{\bf \overline N_3}& r''_I \\
\Omega''_\pm & {\bf det}^{\pm 1} &  {\bf 1} & {\bf 1}&{\bf 1}& r''_\pm  \\
{M^j}_i & {\bf 1} & {\bf  \overline N_1}& {\bf N_2}& {\bf 1}& r_M\\
{\Gamma''^{\,I}}_i & {\bf 1}& {\bf \overline N_1}& {\bf 1}& {\bf N_3}& r''_\Gamma
\end{array}
\ee
\begin{figure}[t]
\begin{center}
\resizebox{0.8\hsize}{!}{
\includegraphics[width=8cm]{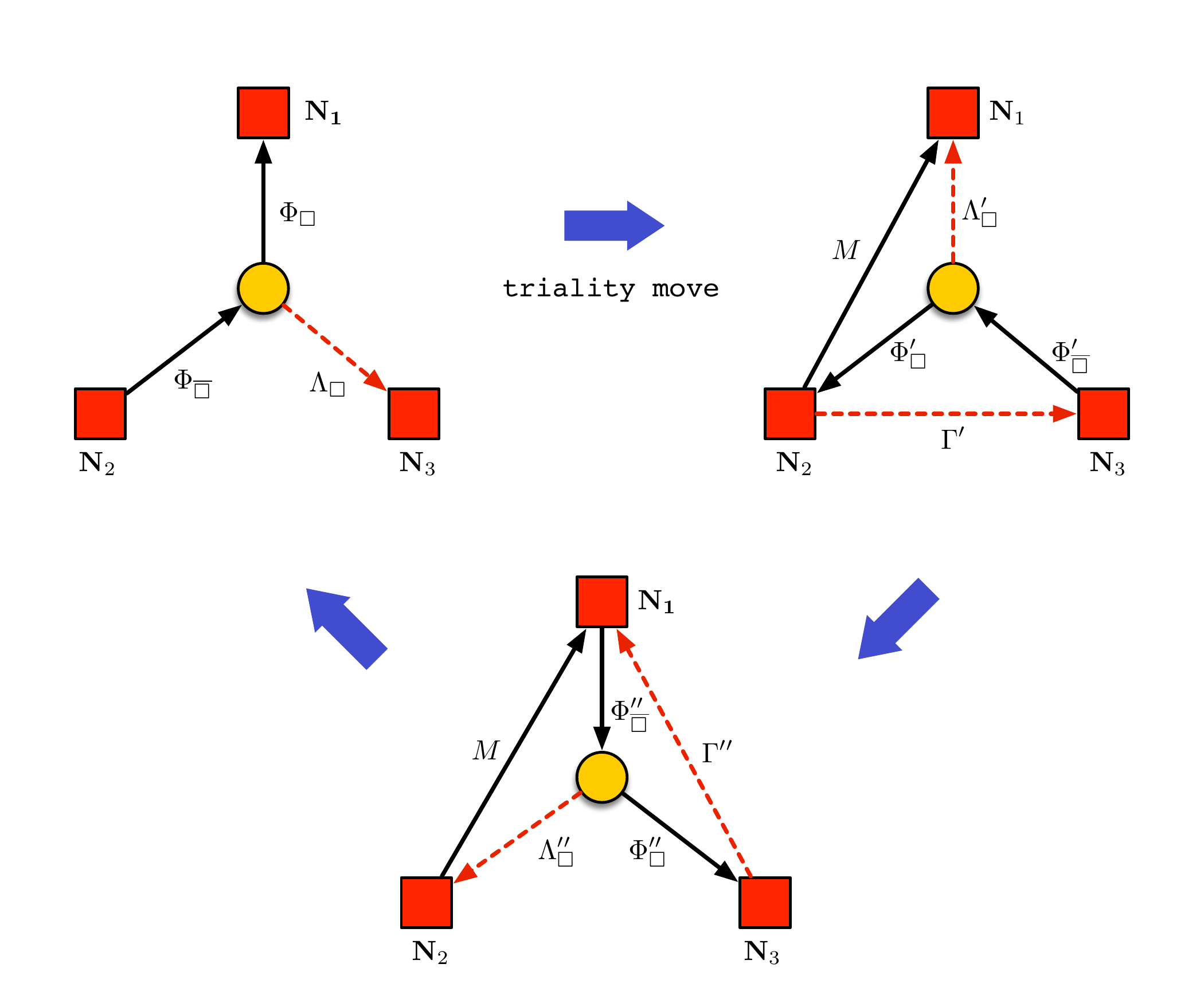}
}
\caption{
The triality cycle for $2d$ $\CN=(0,2)$ SQCD. The fermi fields $\Omega_\pm$ in the determinant representation of $U(N_c)$ are omitted to avoid clutter.
\label{trialitycycle}}
 \end{center}
 \end{figure} 
This theory still contains mesonic gauge-singlet, the chiral multiplets ${M^j}_i$ and another fermi multiplets ${\Gamma''^{\,I}}_i$,  which are given by:
\be
{M^j}_i = {\Phi_{\overline{\tiny\yng(1)}}}^{\,j}\, {\Phi_{{\tiny\yng(1)}}}_{\,i}~, \qquad\quad
{\Gamma''^{\,I}}_i =  {\Phi'_{\overline{\tiny\yng(1)}}}^{\,I}\, {\Lambda'_{{\tiny\yng(1)}}}_{\,i}~,
\ee
in terms of the fields in the first and second theory, respectively.
The fermi multiplets have the non-trivial superpotentials:
\bea\label{inter step2}
&  E_{\Lambda''_j} =0~, \qquad \qquad && J_{\Lambda''_{j}}=  {M^j}_i \, {\Phi''_{\overline{\tiny\yng(1)}}}^{\,i} ~, \cr
& E_{{\Gamma''^{\,I}}_i }=0~, \qquad \qquad && J_{{\Gamma''^{\,I}}_i } =  {\Phi''_{\overline{\tiny\yng(1)}}}^{\,i}\, {\Phi''_{{\tiny\yng(1)}}}_{\,I} ~.
\eea 
Here, we have already integrated out some massive mesons.~\footnote{A direct application  of  \protect\eqref{trialityrule}-\protect\eqref{duality operation} leads to a theory with the mesonic chiral fields $M$ and $M'$, and the mesonic fermi multiplets $\Gamma'$ and $\Gamma''$. We then have:
\bea\nn
& E_{\Lambda''} ={\Phi''_{{\tiny\yng(1)}}} M'~, \qquad &&    E_{\Gamma''}= - M' M~, \qquad && E_{\Gamma'}=0~, \cr
& J_{\Lambda''}= M  {\Phi''_{\overline{\tiny\yng(1)}}}~,   && J_{\Gamma''} =  {\Phi''_{\overline{\tiny\yng(1)}}}{\Phi''_{{\tiny\yng(1)}}}~, && J_{\Gamma'}= M'~.
\eea
The last line states that $M'$ and $\Gamma'$ are massive, and can be integrated out by setting $M'=0$. This leads to \protect\eqref{inter step2}.
}

One can similarly check that a third triality move gives back exactly the original SQCD theory. The duality operation can be conveniently summarized in terms of quiver diagrams, as shown in Figure \ref{trialitycycle}.

\section{Gauged matrix models and sphere compactification}\label{matrix models}
In this section, after reviewing $\CN=1$ matrix models following \cite{Franco:2016tcm}, we explain how one can obtain them from sphere compactification of 2d $\CN=(0,2)$ theories.

\subsection{$\CN=1$ supersymmetric matrix models}
Gauged matrix models (GMM) with $\CN=1$ supersymmetry are matrix models of c-number variables $\varphi$ and Grassmanian variables $\psi$ that transform under a single supersymmetry transformation, which we assume to be nilpotent. By abuse of notation, we will call the integration variables $\varphi$ and $\psi$ the ``bosons'' and ``fermions,'' respectively.

\subsubsection*{Supermultiplets.} Any $\CN=1$ GMM can be defined using the following supersymmetry multiplets:~\footnote{Our definition of a $0d$ fermi multiplet differs slightly from \cite{Franco:2016tcm}. Our present definition is naturally inherited from $(0,2)$ fermi multiplets in two dimensions upon sphere compactification.}

\paragraph{Chiral multiplet.} The $0d$ $\CN=1$ chiral multiplet $\Phi$ has components and supersymmetry transformations:
\be\label{susyPhi}
\Phi = (\phi, \b\phi, \b \psi)~,\qquad \qquad\qquad
\delta \phi=0~, \qquad \delta \b\phi =\b\psi~, \qquad \delta \b\psi=0
\ee
Note that $\delta^2=0$ trivially. Obviously, $\phi$ and $(\b\phi, \b\psi)$ are in fact independent supersymmetry multiplets, but we consider the boson $\phi$ as a complex variables, with $\b\phi$ its complex conjugate---a particular contour of integration on the $\phi$-plane should be part and parcel of the definition of the matrix model. The fermion $\b\psi$ should be considered an independent complex, anti-commuting variable. The superfield $\Phi$ can be charged under some $U(1)$ symmetry, with charge $Q$, which means that the fields $(\phi, \b\phi, \b\psi)$ have charges $(Q, -Q, -Q)$, respectively.

It should be noted that the ``target space'' spanned by the integration variables $\phi$ is a complex space by construction. Moreover, the supersymmetry $\delta$ is naturally identified with the Dolbeault operator $\b \d$ on target space,

\paragraph{Fermi multiplet.} The $0d$ fermi multiplet $\Lambda$ has a single complex fermion $\lambda$, which can be charged. The supersymmetry transformation is determined by:
\be
\delta \lambda= F_\Lambda(\phi)~.
\ee
Here, $F_\Lambda$ is an ``$\CN=1$ superpotential,'' which is an holomorphic function of the bosons $\phi$ in  chiral multiplets. We have $\delta^2 \lambda= 0$  in virtue of \eqref{susyPhi}. The superpotential $F_\Lambda$ must be specified for each fermi multiplet $\Lambda$.

\paragraph{Gaugino multiplet.} The gaugino multiplet $\CV$ is of the form:
\be
\CV= (\chi_0, D_0)~, \qquad \qquad \qquad \delta \chi_0= D_0
\ee
 for some gauge group $\GG$. The ``gaugino'' $\chi_0$ and the real ``auxiliary'' field $D_0$ are valued in the  the adjoint representation of $\Fg= {\rm Lie}(\GG)$.

\subsubsection*{Interaction terms.} Consider a ``linear''  $\CN=1$ matrix model consisting of chiral multiplets $\Phi_i$ and fermi multiplets $\Lambda_I$, with $\phi_i$ valued in $\C$. We have the supersymmetric action:
\be\label{SF def}
S_F = \b F^I(\b \phi) F_I(\phi) + \b \psi^i {\d \b F^I \ov \d \b\phi^i } \lambda_I
\ee
where $F_I$ is the superpotential for $\Lambda_I$.  Incidentally, the action is $\delta$-exact:
\be
S_F = \delta \Big(\b F^I \lambda_I  \Big)~.
\ee
In particular, if $F(\phi)$ is linear in $\phi$, the action \eqref{SF def} is a ``mass term'' and the matrix-model integral is Gaussian.
Another interaction term is available at second order in the fermions:
\be\label{SH def}
S_H = H^{IJ}(\phi) \lambda_I \lambda_J~.
\ee
The holomorphic potential $H^{IJ}= - H^{JI}$ must satisfy the condition $H^{IJ}F_I=0$. It is {\it not} $\delta$-exact.

Next, consider any continuous symmetry group $\GG$ acting on the fields (and commuting with $\delta$), preserved by the potential terms. One can ``gauge'' this symmetry by introducing the corresponding $\Fg$-valued gaugino multiplet with the action:
\be\label{Sg def}
S_{\rm gauge}= S_{D}+ S_{\xi} + S_{\b\phi\phi} = \half D_0^2 - i \xi D_0 + i \b \phi D_0 \phi - i \b \psi \chi_0 \phi~,
\ee
coupling the ``matter fields'' to the gaugino multiplet, with $\xi$ a Fayet-Iliopoulos (FI) term.
These terms are also $\delta$-exact:
\be
 S_{D} = \delta \left( \half D_0\chi_0\right)~, \qquad \quad
  S_{\xi} = \delta \left(-i  \xi \chi_0\right)~, \qquad \quad
S_{\b\phi\phi} = \delta \left( i \b \phi \chi_0\phi \right)~.
\ee
Integrating out $D_0$, we have:
\be
D= -i \left(\b\phi\phi - \xi \phi\right) \equiv -i \mu~,
\ee
where $\mu$ is the moment map for the $\GG$ action on field space. We then obtain:
\be
S_{\rm gauge} \cong \mu^2 - i \b\psi\chi_0 \phi~.
\ee
Since one can tune the coefficient of $S_{\rm gauge}$ at will (at least formally), one finds that the matrix integral receives only contributions from $\mu=0$. In other words, the gauging leaves us with a target space described as a K{\"a}hler quotient of flat space.

\subsubsection*{Anomalies and selection rules.}
Consider a matrix model defined by the schematic integral:
\be\label{defZGMM}
Z = \int  \prod_i d\phi_i \, d \b \phi_i  \, d \b \psi_i \int \prod_I d \lambda_I   \int d\chi_0\, d D_0 \; e^{-S(\phi, \b\phi, \b \psi, \lambda, \chi_0, D_0)}~.
\ee
Here, $\phi, \b \phi$ and $D$ denote complex and real bosons, respectively, while $\psi, \lambda, \chi_0$ are fermions, while the action $S$ is the sum of the terms \eqref{SF def}, \eqref{SH def} and \eqref{Sg def}. In such a model, a symmetry---which leaves $S$ invariant, by definition---can be ``anomalous'' if the integration measure fails to be invariant as well. One can check that this can only happen for an abelian symmetry. Consider a $U(1)$ symmetry that assigns charges $Q_i$ and $Q_I$ to the chiral multiplets $\Phi_i$ and fermi multiplets $\Lambda_I$, respectively. The integration measure has the $U(1)$ charge:
\be\label{0d anomaly gen}
\CA_{U(1)}= \sum_i Q_i - \sum_I Q_I~.
\ee
This anomaly must vanish if $U(1)$ is a {\it gauge} symmetry. For a global symmetry, on the other hand, a non-zero anomaly---that is, a `t Hooft anomaly---implies a selection rule. Consider the observable:
\be
\langle \CO(\Phi, \Lambda) \rangle = \int d\Phi d\Lambda d\CV \; \CO(\Phi, \Lambda)\;  e^{-S}~,
\ee
where the integration measure is the same as in \eqref{defZGMM},  and $\CO$ is a gauge invariant function of the variables. This vanishes whenever $Q[\CO]+ \CA_{U(1)}\neq 0$ \cite{Franco:2016tcm} because Grassmann integration measure cannot be saturated by the expansion of symmetry-invarinat action.

\subsection{$0d$ SQCD and quadrality}
Let us define zero-dimensional $\CN=1$ SQCD as the supersymmetric GMM with $U(N_c)$ gauge symmetry which associates gaugino multiplet coupled to 
$N_1$ chiral multiplets $\Phi_{{\tiny\yng(1)}}$ in the fundamental representation, $N_2$ chiral multiplets $\Phi_{\overline{\tiny\yng(1)}}$ in the anti-fundamental representation, $N_3$  fermi multiplets $\Lambda_{\overline{\tiny\yng(1)}}$   in the anti-fundamental representation, and $N_4$ fermi multiplets $\Lambda_{{\tiny\yng(1)}}$ in the fundamental representation. The fields are summarized in Table \ref{tab: 0dSQCD fields}. In order to cancel the gauge anomaly, we must have:
\be\label{0d AF}
N_1 - N_2 + N_3- N_4=0~.
\ee
All the fermi multiplets have vanishing potentials, $F_I=0$ and $H^{IJ}=0$. 
\begin{table}[t]
\centering
\be\nn
\begin{array}{c|c|cccc}
    & U(N_c)& U(N_1) & U(N_2) & U(N_3) & U(N_4)     \\
\hline
{\Phi_{{\tiny\yng(1)}}}_{\,i} & {\bf N_c} &  {\bf \overline N_1} & {\bf 1}&{\bf 1}&  {\bf 1}\\
{\Phi_{\overline{\tiny\yng(1)}}}^{\,j}  & {\bf \overline N_c} &  {\bf 1} & {\bf N_2}&{\bf 1}&  {\bf 1}\\
{\Lambda_{\overline{\tiny\yng(1)}}}^{\,I} & {\bf \overline N_c} &  {\bf 1} & {\bf 1}&{\bf  N_3}&  {\bf 1} \\
{\Lambda_{{\tiny\yng(1)}}}_{\,J} & {\bf N_c} &  {\bf 1} & {\bf 1}&{\bf 1}&{\bf  \overline N_4} \\
\hline
{{\t\Gamma}^j}_{\;i}& {\bf 1} & {\bf \overline N_1}& {\bf N_2}& {\bf 1}&  {\bf 1}
\end{array}
\ee
\caption{Matter fields in $0d$ $\CN=1$ SQCD. The last line is the gauge-singlet in $0d$ $\t\Gamma$-SQCD. We use $i, j, I, J$ for the $U(N_1)\times U(N_2)\times U(N_3)\times U(N_4)$ flavor indices. }
\label{tab: 0dSQCD fields}
\end{table}

This matrix model is a somewhat ill-defined since the bosonic integral generally diverges, while the fermionic measure is not adequately saturated. Nevertheless, it is interesting to study $0d$ SQCD formally, as a particularly simple starting point. We will consider a  better-behaved model shortly.

\subsection*{$0d$ $\t\Gamma$-SQCD}
\begin{figure}[H]
\begin{center}
\resizebox{0.9\hsize}{!}{
\includegraphics[width=8cm]{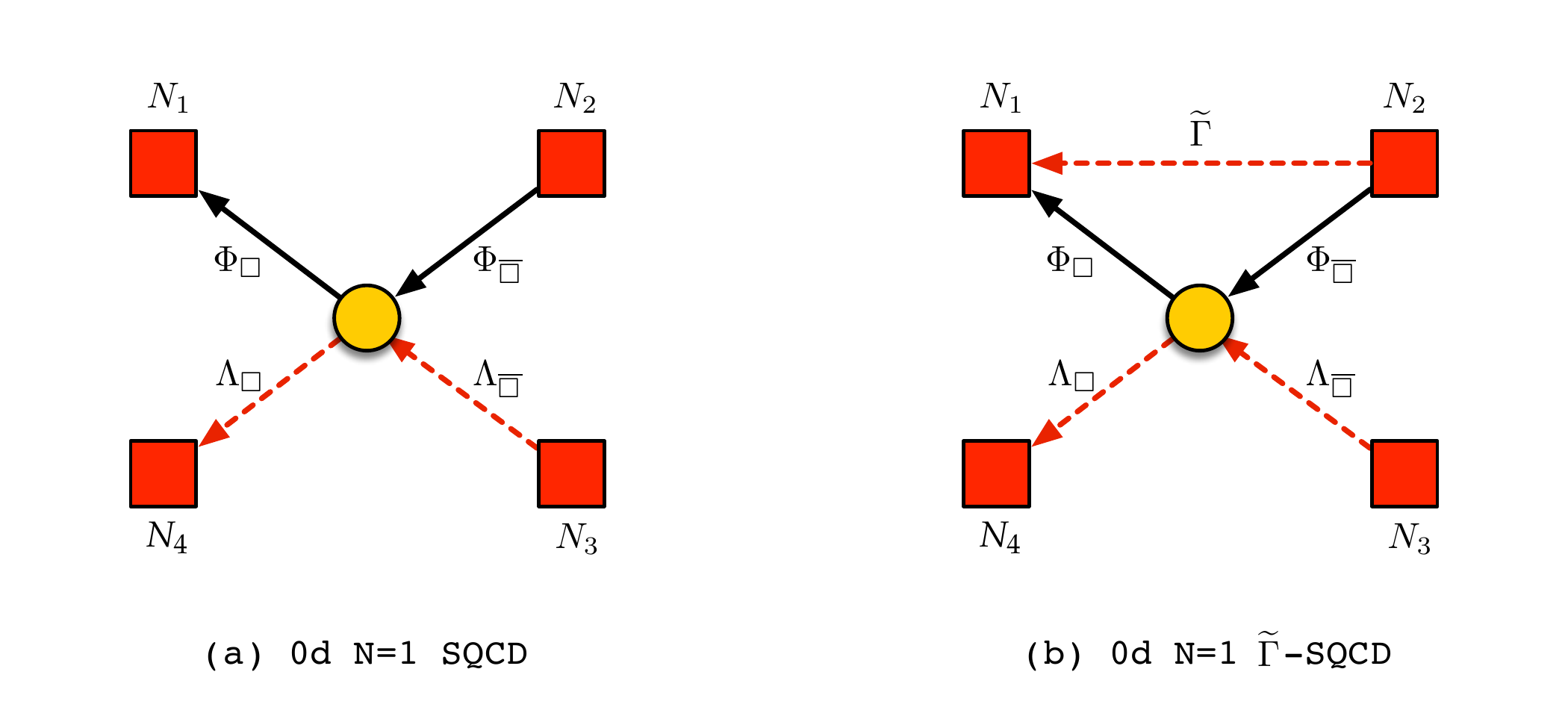}
}
\caption{
Quiver diagrams of two different $0d$ $\CN=1$ SQCD. 
\label{0dsqcd}}
 \end{center}
 \end{figure}

Similarly to $\Gamma$-SQCD in two dimensions, the zero-dimensional $\t \Gamma$-SQCD is defined as a $U(N_c)$ GMM with the (anti-) fundamentally charged fields introduced above, together with additional gauge-singlet fermi fields ${{\t\Gamma}^j}_{\;i}$ with non-trivial $\CN=1$ superpotential:
\be\label{F term tGamma}
F_{{{\t\Gamma}^j}_{\;i}} = {\phi_{\overline{\tiny\yng(1)}}}^{\, j}\, {\phi_{{\tiny\yng(1)}}}_{\, i}~.
\ee
This implies that, in addition to the gauge interactions, the action contains the term:
\be
S_{\t \Gamma}=    {{\overline \phi}_{\overline{\tiny\yng(1)}}}_{\, j}\, {{\overline \phi}_{{\tiny\yng(1)}}}^{\, i}\, {\phi_{\overline{\tiny\yng(1)}}}^{\, j}\, {\phi_{{\tiny\yng(1)}}}_{\, i}
 \;+\; {{\overline \psi}_{\overline{\tiny\yng(1)}}}_{\, j}\, {{\overline \phi}_{{\tiny\yng(1)}}}^{\, i}\; {{\t\Gamma}^j}_{\;i}
 \;-\;    {{\t\Gamma}^j}_{\;i}\;  {{\overline \phi}_{\overline{\tiny\yng(1)}}}_{\, j}\,
 {{\overline \psi}_{{\tiny\yng(1)}}}^{\, i}
 \ee
 The $\phi^4$ term provides a damping factor. Both SQCD and $\t\Gamma$-SQCD are summarized in \fref{0dsqcd}.

\paragraph{Quadrality.} It has been conjectured  that $0d$ SQCD has four equivalent descriptions \cite{Franco:2016tcm}, related by a so-called {quadrality}. The first dual theory is a $U(N_c')$ theory, with dual rank:
\be
N_c'= N_2- N_c~.
\ee
The matter content is given as follows:
\be\label{theory2 0d}
\begin{array}{c|c|cccc}
    & U(N_c)& U(N_1) & U(N_2) & U(N_3)   & U(N_4) \\
\hline
{\Lambda'_{{\tiny\yng(1)}}}_{\,i} & {\bf N_c} &  {\bf \overline N_1} & {\bf 1}&{\bf 1}&{\bf 1} \\
{\Phi'_{{\tiny\yng(1)}}}_{\,j}  & {\bf  N_c} &  {\bf 1} & {\bf \overline N_2}&{\bf 1}& {\bf 1}\\
{\Phi'_{\overline{\tiny\yng(1)}}}^{\,I} & {\bf \overline N_c} &  {\bf 1} & {\bf 1}&{\bf N_3}&{\bf 1}\\
{\Lambda'_{\overline{\tiny\yng(1)}}}^{\,J}  & {\bf \overline N_c} &  {\bf 1} & {\bf 1}&{\bf 1}& {\bf N_4}  \\
{M^j}_i & {\bf 1} & {\bf  \overline N_1}& {\bf N_2}& {\bf 1}&{\bf 1}\\
{{\t\Gamma}'^{\,I}}_{\;\;\;j} & {\bf 1}& {\bf 1}& {\bf \overline N_2}& {\bf N_3}& {\bf 1} \\
{\Xi^j}_J& {\bf 1}& {\bf 1}& {\bf  N_2}& {\bf 1}& {\bf \overline N_4}
\end{array}
\ee
Here the symbols $\Lambda$, $\t\Gamma$ and $\Xi$ denote fermions, while the symbols $\Phi$ and $M$ denote bosons.
The non-vanishing superpotentials are:
\be\label{Fquadra move}
F_{\Lambda'_{\,i} }= {\Phi'_{{\tiny\yng(1)}}}_{\,j} \, {M^j}_i~, \qquad 
F_{{{\t\Gamma}'^{\,I}}_{\;\;\;j}} = {\Phi'_{\overline{\tiny\yng(1)}}}^{\,I}\, {\Phi'_{{\tiny\yng(1)}}}_{\,j}~.
\ee
for ${\Lambda'_{{\tiny\yng(1)}}}_{\,i}$ and ${{\t\Gamma}'^{\,I}}_{\;\;\;j}$, respectively. We also have $H$-terms that  couple ${\Lambda'_{\overline{\tiny\yng(1)}}}^{\,J}$ and ${\Xi^j}_J$, according to:
\be\label{Hquadra move}
H^{ {\Xi^j}_J\, {\Lambda'}^{\,J}}=  {\Phi'_{{\tiny\yng(1)}}}_{\,j}~, \qquad \quad \Rightarrow \qquad \quad
S_H =  {\Phi'_{{\tiny\yng(1)}}}_{\,j} \, {\Xi^j}_J \, {\Lambda'_{\overline{\tiny\yng(1)}}}^{\,J}~.
\ee

\vskip0.2cm
\paragraph{The quadrality move.}  We can think of quadrality as a particular operation on a $0d$ $\CN=1$ quiver, which is locally like the SQCD quiver in Figure~\ref{0dsqcd}. The quadrality move can be summarized as follows: First of all,  the (anti)-fundamental fields are permuted according to:
\be
{\Phi_{\overline{\tiny\yng(1)}}} \;\longrightarrow\; {\Phi'_{{\tiny\yng(1)}}}~, \qquad 
{\Phi_{{\tiny\yng(1)}}} \;\longrightarrow\; {\Lambda'_{{\tiny\yng(1)}}}~, \qquad 
{\Lambda_{{\tiny\yng(1)}}} \;\longrightarrow\; {\Lambda'_{\overline{\tiny\yng(1)}}} ~, \qquad 
{\Lambda_{\overline{\tiny\yng(1)}}} \;\longrightarrow\; {\Phi'_{\overline{\tiny\yng(1)}}}~.
\ee
The gauge group rank transforms according to $N_c \rightarrow N_A- N_c$, where $N_A$ is the number of anti-fundamental chiral multiplets in the original theory. (In the present case, $N_A=N_2$.)
One also introduce the gauge-singlet  $M$, $\t \Gamma$ and $\Xi$, which are identified with ``mesons'' in the original theory:
\be
{M^j}_i = {\Phi_{\overline{\tiny\yng(1)}}}^{\,j}\, {\Phi_{{\tiny\yng(1)}}}_{\,i}~, \qquad\quad
{{\t\Gamma}'^{\,I}}_{\;\;\;j} =  {\Lambda_{\overline{\tiny\yng(1)}}}^{\,I}\, {\overline\phi_{\overline{\tiny\yng(1)}}}_{\,j}~, \qquad\quad
{\Xi^j}_J = {\Phi_{\overline{\tiny\yng(1)}}}^{\,j}\,  {\Lambda_{{\tiny\yng(1)}}}_{\,J}~.
\ee
\begin{figure}[H]
\begin{center}
\resizebox{0.9\hsize}{!}{
\includegraphics[width=8cm]{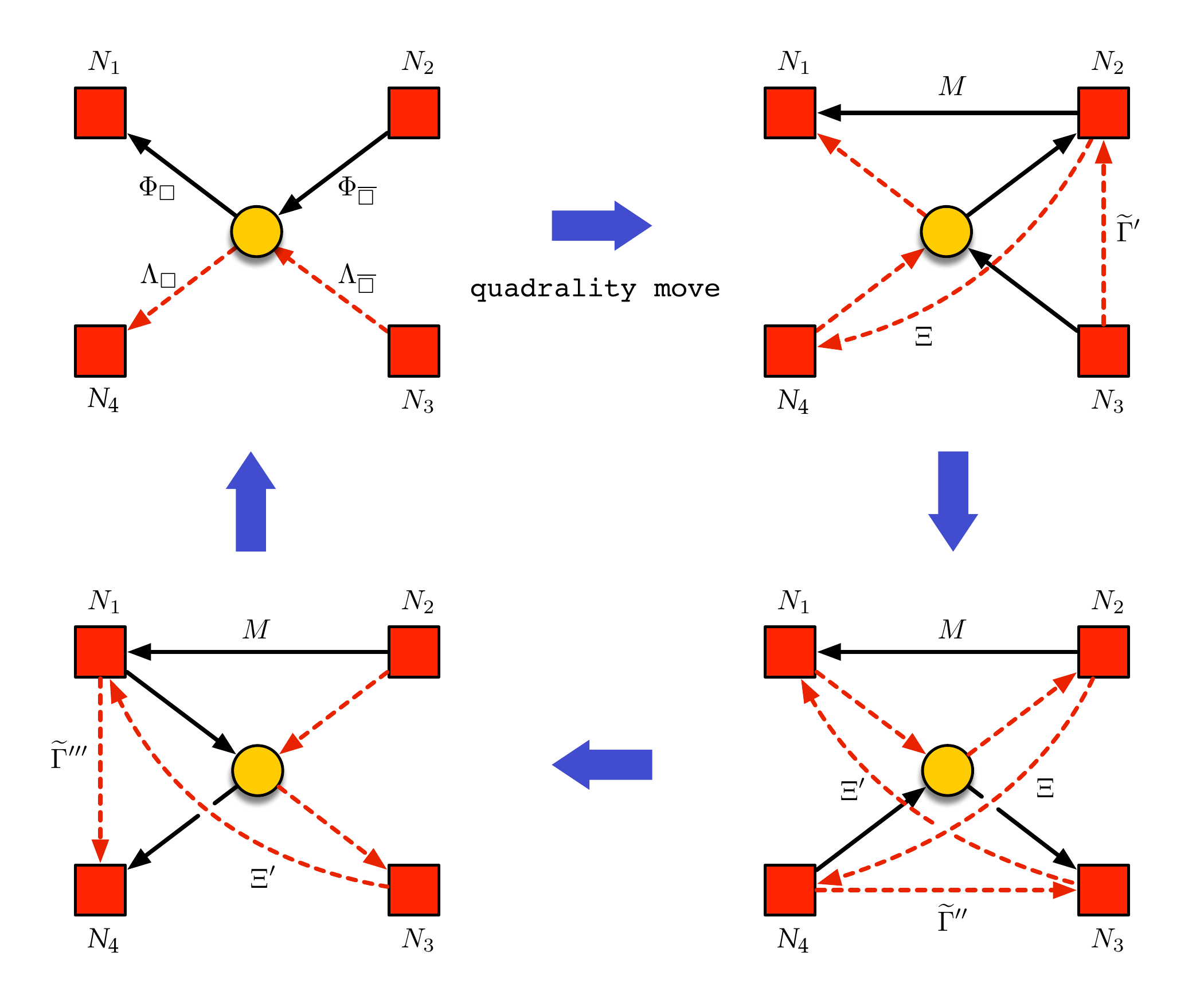}
}
\caption{
The quadrality cycle for $0d$ $\mathcal{N}=1$ SQCD.
\label{quadralitycycle}}
 \end{center}
 \end{figure} 
In the absence of interactions in the original theory, the new interactions are specified by \eqref{Fquadra move} and  \eqref{Hquadra move}. More generally, consider the interaction terms:
\be\label{FH original gen}
F_{\Lambda_J}~, \qquad F_{\Lambda^I}~, \qquad F_\Theta~, \qquad 
H^{\Lambda^I \Lambda_J}~, \qquad
H^{\Lambda_J \Theta}~, \qquad 
H^{\Lambda^I \Theta}~,
\ee
in the original theory, where $\Theta$ denote any other fermi multiplets that are not charged under the $U(N_c)$ gauge theory.   All the superpotential terms \eqref{FH original gen} are holomorphic in ${\Phi_{{\tiny\yng(1)}}}_{\,i}$, $ {\Phi_{\overline{\tiny\yng(1)}}}^{\,j}$ and in any other chiral multiplets $X$ in the larger theory. In the  dual theory obtained after one quadrality move, the new $F$-terms are given by:
\bea\label{dual inter quad 1}
&F_{\Lambda'_i} ={\Phi'_{{\tiny\yng(1)}}}_{\,j} \, {M^j}_i~, \qquad
&& F_{{\Lambda'}^J} = H^{\Lambda^I \Lambda_J}  {\Phi'_{\overline{\tiny\yng(1)}}}^{\,I}~, \cr
&F_{{{\t\Gamma}'^{\,I}}_{\;\;\;j}}  =  {\Phi'_{\overline{\tiny\yng(1)}}}^{\,I}\,   {\Phi'_{{\tiny\yng(1)}}}_{\,j} 
\;- {\d F_{\Lambda^I}\ov \d  {\Phi_{\overline{\tiny\yng(1)}}}^{\,j}}~, \qquad
&& F_{{\Xi^j}_J}=  - {M^j}_i \, {\d F_{\Lambda_J} \ov \d {\Phi_{{\tiny\yng(1)}}}_{\,i}}~.
\eea
For the spectator fermi multiplets $\Theta$, the potentials $F_{\Theta}$ are the same as in the original theory after substituting the gauge-invariant combination $\Phi^j \Phi_i$ with the ``meson'' singlet ${M^j}_i$.
The non-zero $H$-term potentials are given by:~\footnote{This is given up to some signs, which can be fixed for consistency with $\Tr(H^{IJ} F_I)=0$ in any given example.}
\bea\label{dual inter quad 2}
&H^{{\Lambda^{' J}} {\Lambda'}_i} =  {\d F_{\Lambda_J} \ov \d {\Phi_{{\tiny\yng(1)}}}_{\,i}}~, \quad 
&& H^{{\Xi^j}_J {\Lambda^{' J}}} ={\Phi'_{{\tiny\yng(1)}}}_{\,j}~, \quad 
&& H^{{{\t\Gamma}'^{\,I}}_{\;\;\;j}\, {\Xi^j}_J } = -H^{\Lambda^I \Lambda_J}~, \cr
& H^{ {\Lambda'}_i \, \Theta}  = {\d H^{\Lambda^I \Theta}\ov \d {\Phi_{{\tiny\yng(1)}}}_{\,i}}\, {\Phi'_{\overline{\tiny\yng(1)}}}^{\,I}~, \quad
&& H^{ {\Xi^j}_J\, \Theta}  = {\d H^{\Lambda_J \Theta} \ov \d  {\Phi_{\overline{\tiny\yng(1)}}}^{\,j}}~, \quad
&& H^{ {{\t\Gamma}'^{\,I}}_{\;\;\;j}\, \Theta}  = -{M^j}_i  \, {\d H^{\Lambda_J \Theta} \ov \d  {\Phi_{\overline{\tiny\yng(1)}}}^{\,j}}~.
\eea
These transformations rules for the interactions were first discussed in \cite{Franco:2017lpa}.

\paragraph{The quadrality cycle.} Using these rules, it is easy to apply the quadrality move repeatedly.
After four quadrality moves, one recovers the original theory. The field content can be conveniently summarized in quiver notation, as depicted in Figure \ref{quadralitycycle}. (The interactions are essentially given by all the allowed cycles in each quiver. Two-cycles are mass terms, and the relevant fields should be ``integrated out.'') 
The gauge group rank transforms as:
\be
N_c'= N_2-N_c~, \quad N_c''= N_3- N_c'~, \quad N_c'''= N_4-N_c''~, \quad N_c''''= N_1- N_c'''=N_c~,
\ee
where the last equation follows from \eqref{0d AF}. One can also check that the four matrix models have the same 't Hooft anomalies \cite{Franco:2016tcm}. The quadrality cycle for $\t\Gamma$-SQCD can be constructed similarly---its quadrality cycle is almost identical to Figure \ref{quadralitycycle}, except that there are no chiral multiplet mesons $M$ at any step.

\subsection{Sphere compactification of $2d$ SQCD \label{scompsp}}
Consider an $\CN=(0,2)$ gauge theory on a sphere, $S^2$, with the half-twist, as studied {\it e.g.} in \cite{Closset:2015ohf}. We are interested in what happens when we send the radius of $S^2$ to zero. On general ground, one expect that only the zero-modes survive. For a neutral scalar field, that would just be the $s$-wave; more generally, any field has a zero-modes if it is valued in some holomorphic vector bundle over $S^2$ which admits holomorphic sections.

The counting of zero-modes goes as follows \cite{Closset:2015ohf}. Consider first the 2d $\CN=(0,2)$ vector multiplet. There exists a supersymmetric configuration for any choice of integer-quantized gauge fluxes $\m$ through $S^2$:
\be
{1\ov 2\pi} \int_{S^2} da = \m~,
\ee
where $a_\mu$ is the $2d$ gauge field and $\m$ is valued in $\Gamma_{\GG^\vee}$, the magnetic flux lattice of $\GG$. Let $e^a$ ($a=1, \cdots, {\rm rank}(\GG)$) denote a basis of $\Gamma_{\GG^\vee}$ such that $\rho(e^a)\in \Z$ for any weight $\rho$. Consider $\GG= U(N_c)$ for simplicity. The gauge flux $\m = \m_a e^a$, with:
\be\label{gauge flux m 2d}
(\m_a)= (\m_{(1)}, \cdots, \m_{(1)}, \m_{(2)}, \cdots, \m_{(2)}, \cdots, \m_{(s)}, \cdots, \m_{(s)})~, \qquad \m_{(l)}\in \Z~,
\ee
breaks $U(N_c)$ to the Levi subgroup $\prod_{l=1}^s U(n_l)$, with $\sum_{l=1}^s n_l = N_c$.

Upon reduction on the sphere, the gaugino has a zero-mode, leading to a $0d$ gaugino multiplet:~\footnote{Here we have set to one the dimensionful parameter $\sqrt{{\rm vol}(S^2)/e^2}$, with $e^2$ the 2d gauge coupling.}
\be
\t \chi^{(2d)} = \chi_0~, \qquad  D^{(2d)}-2 i f_{1\b 1}=  i D_0~.
\ee
In the presence of the gauge flux \eqref{gauge flux m 2d}, we obtain a matrix model with gauge group $\prod_k U(n_l)$. The 2d ``W-bosons'' become $0d$ fermi multiplets in bifundamental representations (connecting different $U(n_l)$ factors). In the zero-flux sector, $\m=0$, we simply obtain a $0d$ matrix model with gauge group $U(N_c)$. This is the sector we will focus on.

The zero-modes of the 2d matter fields are similarly accounted for. We refer to Appendix \ref{convention} for more detail. Given a $2d$ $\CN=(0,2)$ {\it chiral multiplet} $\Phi_i$ in the representation $\FR_i$ of $\Fg$, with $\rho_i \in \FR$ the weights of the representation, and with $R$-charge $r_i$, let us define the integer:
\be
\n_{\rho_i} = \rho_i(\m) - r_i +1~.
\ee
for each field component $\Phi_{\rho_i}$. Upon reduction on $S^2$, we obtain a number of $0d$ chiral and/or fermi multiplets, depending on the sign of $\n_{\rho_i}$:
\be\label{zero modes count chiral}
\Phi_{\rho_i} \rightarrow \begin{cases}
\n_{\rho_i} \quad\quad\;\; \text{0d chiral multiplets}\; \Phi_{\rho_i} &\; {\rm if}\;\;  \,\n_{\rho_i} \geq 0~,\cr
-\n_{\rho_i}   \quad\quad  \text{0d fermi multiplets} \; \Lambda_{\rho_i}  &\; {\rm if}\;\;\; \n_{\rho_i} < 0~.\end{cases}
\ee
In particular, in the zero-flux sector, the $2d$ chiral multiplet  $\Phi_i$ gives rise to $0d$ chiral or a fermi multiplets in the same representation $\FR_i$ of $\GG$ depending on whether $r_i <1$ or $r_i>1$, respectively.

Given a $2d$ $\CN=(0,2)$ {\it fermi multiplet} $\Lambda_I$ in the representation $\FR_I$ of $\Fg$ and with $R$-charge $r_I$, let us define the integer:
\be
\n_{\rho_I} = \rho_i(\m) - r_I~.
\ee
for each field component $\Lambda_{\rho_I}$. Upon reduction, we have:
\be\label{zero modes count Fermi}
\Lambda_{\rho_I} \rightarrow \begin{cases}
\n_{\rho_I} \quad\quad \,\;\text{0d fermi multiplets}\; \Lambda_{\rho_I} &\; {\rm if}\;\;  \n_{\rho_I} \geq 0~,\cr
-\n_{\rho_I} \quad\quad \text{0d fermi multiplets}\; \b\Lambda_{\b \rho_I} &\; {\rm if}\;\; \, \n_{\rho_I} < 0~,
\end{cases}
\ee
where $\rho_I \in \b\FR_I$ denote the weights of the conjugate representation. In particular, in the zero-flux sector, a $2d$ fermi multiplet in a representation $\FR$ gives rise to $0d$ fermi multiplets in the representation $\FR$ if $r_I <0$, or in the conjugate representation $\b\FR$ if $r_I>0$.

The 2d holomorphic potentials $E$ and $J$ give rise to the $0d$ holomorphic potentials $F$ and $H$. Let us denote by $(\phi_i, \psi_i, \b\phi^i, \b \psi^i)$ and by $(\lambda_I, \b\lambda^I)$ the $2d$ fields in chiral and fermi multiplets, respectively, with superpotentials $E_I= E_I(\phi)$ and $J^I= J^I(\phi)$ such that $\Tr(J^I E_I)=0$.  The $2d$ interactions are given by:
\be
{\mathcal L}_{\rm pot}= \b J_I J^I + \b E^I E_I+ \left(\b \psi^i {\d \b E^I \ov \d \b \phi^i} + {\d J^I \ov \d \phi_i} \psi_i \right)\lambda_I  - \left(\psi_i  {\d E_I \ov \d \phi_i}+ {\d \b J_I \ov \d \b\phi^i} \b \psi^i\right)\b \lambda^I~,
\ee
from which one can read off the interaction terms in the matrix model. Let us expand the $2d$ indices $i, I$ into:
\be
i \rightarrow (j, K)~, \qquad I \rightarrow (M, N)~,
\ee
where the righ-hand-side index $j$ runs over the $0d$ chiral multiplets, and the right-hand-side indices  $K, M, N$ run over the $0d$ fermi multiplets, obtained according to \eqref{zero modes count chiral} and \eqref{zero modes count Fermi}.~\footnote{That is, the $0d$ fermions $\lambda_K$ come from the second line in \protect\eqref{zero modes count chiral}, and the $0d$ fermions $\lambda_M$ and $\lambda_N$ come from the first and second line of \protect\eqref{zero modes count Fermi}, respectively.} In two dimensions, we have $\delta \Lambda_I= E_I$ and $\delta \b\Lambda^I = J^I$ (on-shell) under the one supersymmetry that survives the half-topological twist.  Therefore, to obtain a consistent reduction to $0d$, we need that:
\be\label{red condition}
J^M\big|_{S^2}=0~, \qquad E_N\big|_{S^2}=0~.
\ee
In that case, we have the $0d$ fermi multiplets given in terms of the $2d$ zero-modes by: 
\be
\lambda_K = \psi^K~,\qquad \lambda_M = \lambda_M~,\qquad \lambda_N= \b \lambda^N~,
\ee
 with the $0d$ holomorphic potentials:
\be\label{F H from E J}
F_K=0~, \quad F_M = E_M~, \quad F_N= J^N~, \quad H^{MK}= {\d J^M \ov \d \phi_K}~, \quad H^{NK}=  {\d E_N \ov \d \phi_K}~.
\ee
In particular, the supersymmetric condition $F_M H^{MK}+ F_N H^{NK}=0$ follows from the 2d condition $J^I E_I=0$.

Finally, let us note that the cancellation of the $U(1)$ gauge anomalies for the $0d$ GGM follows from the cancellation of the mixed anomaly between the $U(1)$'s and the $R$-symmetry in two dimensions:
\be\label{RQ AF}
 \sum_i  Q_i (r_i-1) - \sum_I Q_I r_I= 0~,
\ee
where $i$ and $I$ run over the 2d chiral and fermi multiplets, respectively, as above. The condition \eqref{RQ AF} is necessary for the $2d$ $R$-symmetry to exist, and thus for the half-twist to exist. Given the $0d$ fields obtained upon reduction on $S^2$, one easily checks that \eqref{RQ AF} implies that the $0d$ gauge anomalies \eqref{0d anomaly gen} vanish. This can be also understood, more generally, in terms of the reduction of the anomaly polynomial of the $2d$ theory on $S^2$ with the $U(1)_R$ flux. If $F_R$ and $F_G$ denote the $U(1)_R$ and $U(1)$ field strengths in $2d$, the relevant terms are:
\begin{align}
\int_{S^2 \times \R^2} \text{tr} (F_R F_G) = -2 \pi \int_{\R^2} \text{tr} (F_G)~,
\end{align}
so that any $\Tr(RQ)$ quadratic anomaly in $2d$ reduces to the linear $\Tr(Q)$ anomaly in $0d$.
See for instance \cite{Gadde:2015wta,Bobev:2017uzs} for further discussions in similar contexts.

\subsection{Quadrality from triality \label{quadrality from triality}}
Using the above rules, it is straightforward to study the dimensional reduction of $2d$ SQCD to $0d$. We focus on the zero-flux sector. Consider 2d SQCD with the $R$-charges:
\be\label{Rcharges 1}
\begin{array}{c|cccccc}
    & {\Phi_{{\tiny\yng(1)}}}_{\,i}&{\Phi_{\overline{\tiny\yng(1)}}}^{\,j} & {\Lambda_{{\tiny\yng(1)}}}_{\,I} & {\Lambda_{{\tiny\yng(1)}}}_{\,J}& {\Lambda_{{\tiny\yng(1)}}}_{\,L}  &\Omega_\pm  \\
\hline
U(1)_R & 0 & 0 & 1 & -1 & 0  &0
\end{array}
\ee
\begin{figure}[t]
\begin{center}
\resizebox{0.9\hsize}{!}{
\includegraphics[width=8cm]{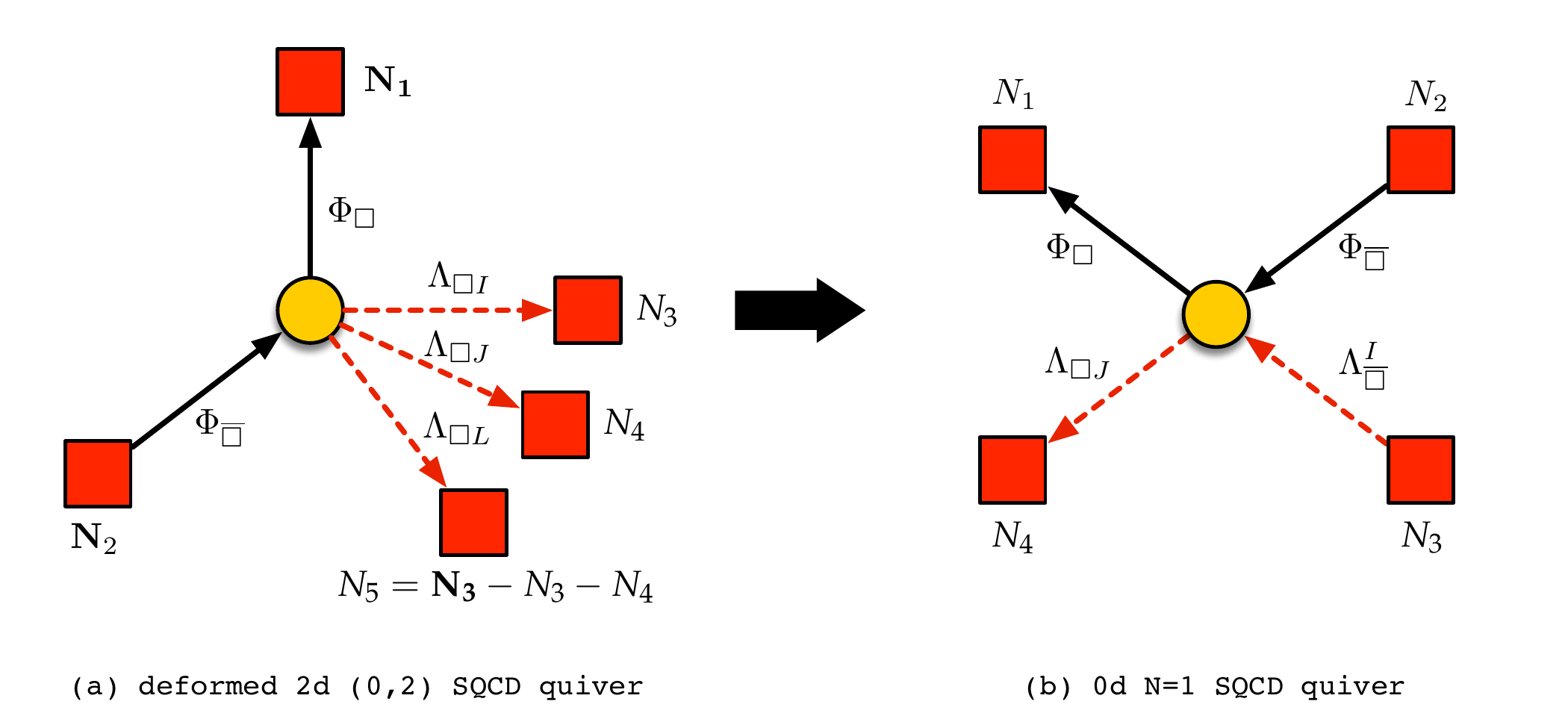}
}
\caption{
The reduction with the half-topological twist from (a) deformed $2d$ $\mathcal{N}=(0,2)$ SQCD to (b) $0d$ $\mathcal{N}=1$ SQCD.
\label{reductionsqcd}}
 \end{center}
 \end{figure} 
This choice of $R$-charges breaks down the $U({\bf N}_3)$ factor of the $2d$ flavor group to:
\be
U({\bf N}_3) \rightarrow  U(N_3) \times U(N_4) \times U(N_5)~, \qquad\qquad N_5 \equiv {\bf N}_3- N_3-N_4~,
\ee
where we labelled the indices $I\rightarrow (I, J, L)$ with $I=1, \cdots N_3$, $J= 1, \cdots, N_4$ and $L= 1, \cdots, N_5$. We also take $N_1= {\bf N}_1$ and $N_2= {\bf N}_2$, with indices $i$ and $j$, respectively. By restriction to the zero-modes, we exactly obtain $0d$ SQCD with the field content of Table \ref{tab: 0dSQCD fields}.

If we consider $2d$ $\Gamma$-SQCD instead, we must choose the $R$-charge assignment:
\be
r_{{\Gamma^i}_j}= 1~,
\ee
for the gauge-singlet fermi multiplets ${\Gamma^i}_j$, consistently with the $J$-potential \eqref{superpot GammaSQCD}. This directly leads to $0d$ $\t \Gamma$-SQCD. The potential term \eqref{F term tGamma} follows from \eqref{superpot GammaSQCD} together with \eqref{F H from E J}---we simply have $F_{\t\Gamma}= J_\Gamma$ in this case. 



The analysis of the other $2d$ theories related to SQCD by triality is similar. Consider the theory \eqref{2d SQCD dual1} obtained after one triality move. The $R$-charge assignment dual to \eqref{Rcharges 1} reads:
\be\label{Rcharges 2}
\begin{array}{c|cccccccccc}
    & {\Lambda'_{{\tiny\yng(1)}}}_{\,i}&{\Phi'_{{\tiny\yng(1)}}}_{\,j}& {\Phi'_{\overline{\tiny\yng(1)}}}^{\,I}  & {\Phi'_{\overline{\tiny\yng(1)}}}^{\,J}& {\Phi'_{\overline{\tiny\yng(1)}}}^{\,K} &{M^j}_i &{\Gamma'^{\,j}}_I&{\Gamma'^{\,j}}_J&{\Gamma'^{\,j}}_L  \\
\hline
U(1)_R & -1 & 0 & 0 & 2 & 1  & 0&1 & -1 & 0
\end{array}
\ee
Upon reduction, this gives exactly the matrix model obtained from $0d$ SQCD after one quadrality move, with matter content \eqref{theory2 0d} and the interaction terms  \eqref{Fquadra move}-\eqref{Hquadra move}. Moreover, $0d$ t'Hooft anomaly matching directly follow from the matching of the $2d$ anomalies under triality.

\begin{figure}[t]
\begin{center}
\resizebox{0.9\hsize}{!}{
\includegraphics[width=8cm]{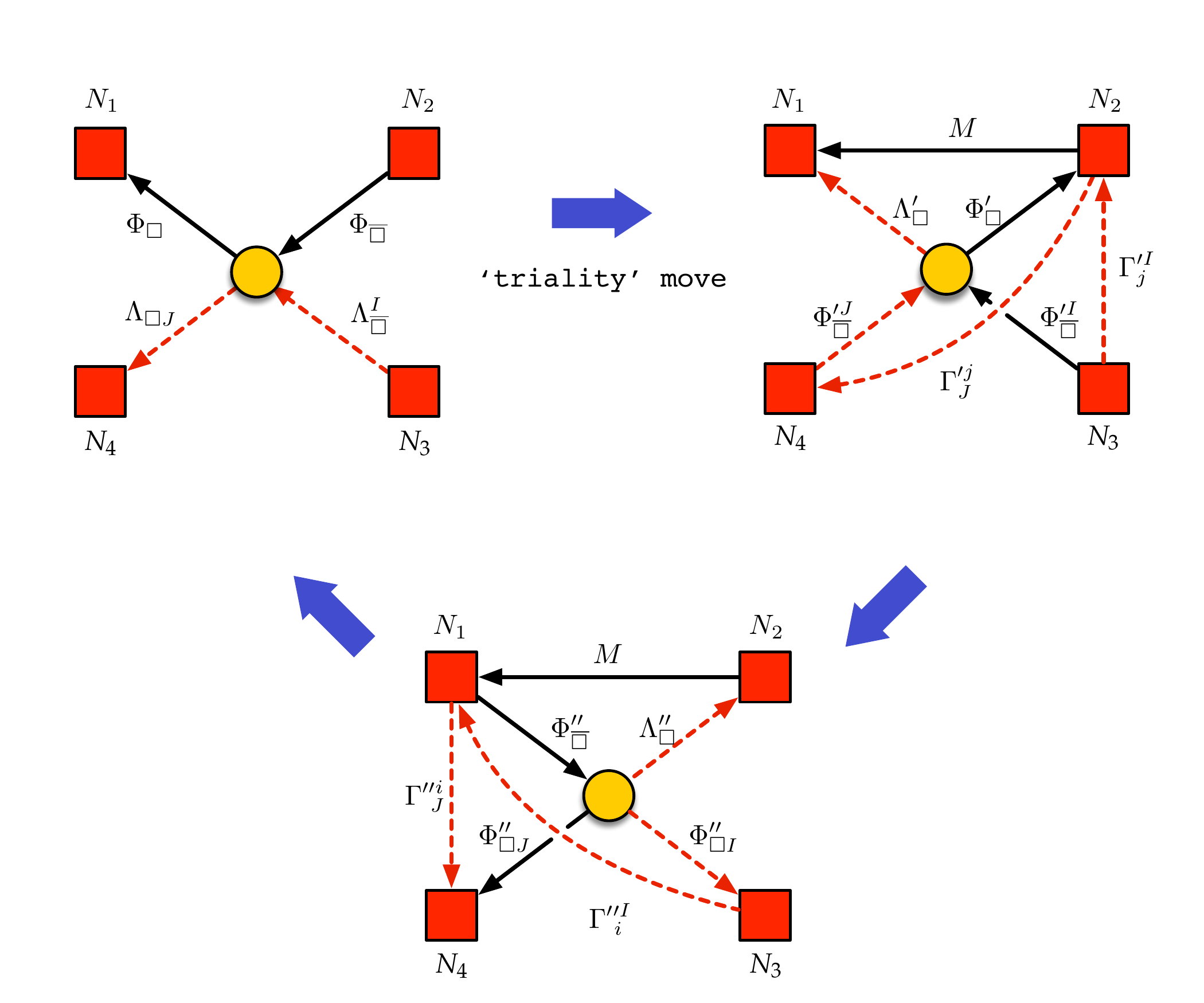}
}
\caption{
The triality (order-3) cycle of $0d$ $\mathcal{N}=1$ SQCD. From the first move interpreted as acting on the $0d$ theory, one can recover the full quadrality cycle of Figure \protect\ref{quadralitycycle}.
\label{0dtriality}}
 \end{center}
 \end{figure}

We can similarly consider the third theory in the triality chain in $2d$. The corresponding $R$-charge assignment reads:
\be\label{Rcharges 3}
\begin{array}{c|cccccccccc}
    & {\Phi''_{\overline{\tiny\yng(1)}}}^{\,i}&{\Lambda''_{{\tiny\yng(1)}}}_{\,j}& {\Phi''_{\tiny\yng(1)}}_{\,I}  & {\Phi''_{\tiny\yng(1)}}_{\,J}& {\Phi''_{\tiny\yng(1)}}_{\,K} &{M^j}_i &{\Gamma''^{\,I}}_i&{\Gamma''^{\,J}}_i&{\Gamma''^{\,L}}_i  \\
\hline
U(1)_R & 0 & -1 & 2 & 0 & 1  & 0& -1 & 1 & 0
\end{array}
\ee
One can check that this exactly recovers {\it three} out of the {\it four} $0d$ theories in the quadrality cycle. This is summarized in \fref{0dtriality}.

From this discussion, we see that we recover most the quadrality, but we are still missing the lower-right corner in Figure~\ref{quadralitycycle}. However, if we think about the triality and quadrality ``moves'' as particular operations on a $U(N)$ gauge theory, as described above, we can simply recover one single quadrality move from one triality move. 
Indeed, consider 2d SQCD with the $R$-charge assignment \eqref{Rcharges 1} and generic interactions such that \eqref{red condition} holds. Upon reduction, we obtain $0d$ SQCD with interaction terms:
\be
F_I = J^I~, \qquad F_J = E_J~, \qquad H^{II}=H^{JJ}=H^{IJ}=0~.
\ee
A triality move gives rise to a new theory with the new 2d interactions \eqref{duality operation}. Reducing this dual theory to $0d$ with the dual $R$-charge assignment \eqref{Rcharges 2} and using \eqref{F H from E J}, one can check that the new interactions in the $0d$ theory are exactly given by \eqref{dual inter quad 1}-\eqref{dual inter quad 2}, in the special case where $H=0$ in the original theory.
This gives a simple derivation of the $0d$ quadrality move from the $2d$ triality move, including all interaction terms. One can then check that the quadrality move is indeed an operation of order {\it four} on the $0d$ theory. This completes the derivation of quadrality from triality.

\subsection{Comment on the non-zero flux sectors}
We have seen that $0d$ SQCD, a $U(N_c)$ gauged matrix model, describes the zero-mode sector of $2d$ $\CN=(0,2)$ SQCD  on a half-twisted sphere, in the absence of gauge fluxes---or, for that matter, of any background flux for the flavor symmetries.~\footnote{Or, we may think of the $R$-charge assignments as being the results of certain flavor fluxes, that explicitly break the $2d$ flavor group down to the $0d$ flavor group as specified above.}  
On the other hand,  $2d$ observables on the sphere are expected to receive contributions from an infinite number of flux sectors, in general. Using the rules above, one can easily write down the  matrix model that one obtains by dimensional reduction in the presence of some particular gauge flux \eqref{gauge flux m 2d}. It is essentially a quiver GMM with gauge group $\prod_l U(n_l)$, with bifundamental fermi multiplets connecting the nodes (coming from the $2d$ $W$-bosons). In addition, the $2d$ fields $\Omega_\pm$ give rise to $0d$ fermions that couple all the $U(n_l)$ gauge factors together. It might be interesting to study those more complicated matrix models systematically.

\subsection{Beyond SQCD: a comment on brane hyper-brick models}

It is interesting to ask how the reduction on a sphere with half-topological twist of a $2d$ $(0,2)$ theory might look like when the $2d$ theory is not a SQCD-type theory, but rather a worldvolume theory of D1-branes at a tip of a Calabi-Yau fourfold singularity. 
Such theories have been extensively studied in the context of brane brick models \cite{Franco:2015tna, Franco:2015tya} and were extended to worldvolume theories of D$(-1)$-branes at Calabi-Yau fivefold singularities \cite{Franco:2016tcm} known as brane hyper-brick models. 
Here we would like to point out that the ``orbifold reduction'' of  \cite{Franco:2016fxm} has features similar to the twisted $S^2$ reduction. It would be interesting to study this further.

Reductions from $4d$ $\mathcal{N}=1$ theories to brane brick models, from the point of view of the underlying brane configurations, were introduced in \cite{Franco:2016fxm}.
These $4d$ $\mathcal{N}=1$ theories are worldvolume theories on D3-branes probing toric Calabi-Yau threefold singularities and are realized in terms of type IIB brane configurations known as brane tilings \cite{Franco:2005rj}.
Simple dimensional reduction of a brane tiling corresponding to a toric Calabi-Yau threefold CY${}_3$ leads to a $2d$ $\CN=(2,2)$ theory corresponding to a Calabi-Yau fourfold of the form CY${}_{3}\times \mathbb{C}$. 
When expressed in terms of $2d$ $\CN=(0,2)$ multiplets, the $2d$ $\CN=(2,2)$ theory can be represented by a brane brick model with $2d$ $\CN=(0,2)$ adjoint chiral multiplets originating from the $4d$ $\mathcal{N}=1$ vector multiplets under dimensional reduction.

 \begin{table}[h!!!]
  \centering
  \begin{tabular}{c|c|c|c}
    & Brane Configuration & T-Duality & D-Brane Probe \\ \hline 
    (a)
    &
    \begin{tabular}{c|cccccccccc}
      \; & 0 & 1 & 2 & 3 & 4 & 5 & 6 & 7 & 8 & 9 
      \\
      \hline
      D5 & $\times$ & $\times$ & $\times$ & $\times$ &  $\cdot$ &  $\times$ &  $\cdot$ &  $\times$   &  $\cdot$ &  $\cdot$
      \\
      NS5 & $\times$ & $\times$ & $\times$ & $\times$ & \multicolumn{4}{c}{----- \ $\Sigma$ \ -----} &  $\cdot$ &  $\cdot$
    \end{tabular}
    &
    $\stackrel{\mbox{2 times}}{\longleftrightarrow}$
    &
    D3 $\perp$ CY3  
    \\ \hline
    (b)
    &
    \hspace{-0.23cm}
    \begin{tabular}{c|cccccccccc}
      \; & 0 & 1 & 2 & 3 & 4 & 5 & 6 & 7 & 8 & 9 
      \\
      \hline
      D4 & $\times$ & $\times$ & $\cdot$ & $\times$ &  $\cdot$ &  $\times$ &  $\cdot$ &  $\times$   &  $\cdot$ &  $\cdot$
      \\
      NS5 & $\times$ & $\times$ & \multicolumn{6}{c}{----------- \ $\Sigma$ \ -----------} &  $\cdot$ &  $\cdot$
    \end{tabular}
    &
    $\stackrel{\mbox{3 times}}{\longleftrightarrow}$
    &
    D1 $\perp$ CY4  
    \\ \hline
    (c)
    &
        \hspace{-0.38cm}
    \begin{tabular}{c|cccccccccc}
      \; & 0 & 1 & 2 & 3 & 4 & 5 & 6 & 7 & 8 & 9 
      \\
      \hline
      D3 & $\cdot$ & $\times$ & $\cdot$ & $\times$ &  $\cdot$ &  $\times$ &  $\cdot$ &  $\times$   &  $\cdot$ &  $\cdot$
      \\
      NS5 & \multicolumn{8}{c}{--------------- \ $\Sigma$ \ ---------------} &  $\cdot$ &  $\cdot$
    \end{tabular}
    &
    $\stackrel{\mbox{4 times}}{\longleftrightarrow}$
    &
    D$(-1)$ $\perp$ CY5  
    \\
  \end{tabular}
  \caption{
        The various brane configurations for supersymmetric gauge theories arising on D-brane probes at toric Calabi-Yau singularities. 
        (a) Brane tilings give rise to $4d$ $\mathcal{N}=1$ supersymmetric gauge theories as worldvolume theories D3-branes at toric Calabi-Yau threefold singularities, (b) brane brick models give rise to $2d$ $\CN=(0,2)$ theories as worldvolume theories of D1-branes at toric Calabi-Yau fourfold singularities, and (c) brane hyper-brick models give rise to $0d$ $\mathcal{N}=1$ gauged matrix models as worldvolume theories of D$(-1)$-branes at toric Calabi-Yau fivefold singularities.
  }
\label{t:branes}
\end{table}
 
When the dimensionally reduced theory is abelian, in which case the probed Calabi-Yau becomes the classical moduli space of the corresponding supersymmetric gauge theory, the adjoint chiral multiplets coming from the reduction parameterize the $\mathbb{C}$ factor of CY${}_{3}\times \mathbb{C}$.
Brane tilings are realized in terms of D5-branes suspended between a NS5-brane that wraps a holomorphic surface $\Sigma$, which can be mapped to a bipartite periodic graph on a 2-torus $T^2$.~\footnote{This bipartite graph is also known in the literature as a dimer model. }
Similarly, brane brick models are type IIA brane configurations of D4-branes suspended between a NS5-brane that wraps a complex 2-dimensional holomorphic surface $\Sigma$.
They can be represented by periodic tessellations of a 3-torus $T^3$.
When a brane tiling is dimensionally reduced, the resulting brane brick model can be constructed using layers of the original brane tiling that wrap around an additional $S^1$ cycle parameterized by the multiplets coming from the dimensional reduction, in particular the adjoint chiral multiplet parameterizing the $\mathbb{C}$ factor of the moduli space \cite{Franco:2015tya}.
This additional $S^1$ cycle gives rise to the full $T^3$ on which the brane brick model is defined.

In order to break supersymmetry down to $(0,2)$, an abelian orbifold was introduced in \cite{Franco:2016fxm} that has the geometric effect of mixing the $\mathbb{C}$ factor with the rest of the Calabi-Yau space. 
For example, a $\Z_2$ orbifold  has the effect of doubling the number of gauge groups in the $2d$ theory, with adjoint chiral multiplets from the original dimensional reduction becoming pairs of bifundamental chiral mutliplets.
These pairs of bifundamental chiral multiplets transform in conjugate representations of the two gauge groups that come from the original single gauge group under the $\mathbb{Z}_2$ orbifold.
Alternatively, one can introduce a reflection symmetry in addition to the $\mathbb{Z}_2$ orbifold, which has the effect of mapping the adjoint chiral mutliplets into a pair of bifundamental chiral multiplets transforming in the same representation rather than in conjugate representations.

The same orbifold reduction (that is, combining dimensonal reduction and orbifolding) can also be used to obtain $0d$ $\mathcal{N}=1$ matrix models from brane brick models.
The $0d$ $\mathcal{N}=1$ matrix models one obtains from orbifold reduction are worldvolume theories of D$(-1)$-branes at toric Calabi-Yau fivefold singularities.
These theories are called ``brane hyper-brick models'' \cite{Franco:2016tcm} and are realized in terms of a type IIB brane configuration of D3-branes suspended between a NS5-brane wrapping a 3-complex dimensional holomorphic surface $\Sigma$. 
The brane configuration can be represented in terms of a tessellation of a 4-torus $T^4$ that originates from a complex coordinate (tropical) projection of $\Sigma$. 
Under orbifold reduction, these brane hyper-brick models on $T^4$ can be built from copies of brane brick models on $T^3$. 
The additional $S^1$ cycle is parameterized by pairs of bifundamental chiral mutliplets that either are in the same or conjugate representations of pairs of distinct gauge groups originating from the orbifolding.
The map due to orbifold reduction from adjoint multiplets charged under a single gauge group to a pair of bifundamental chiral multiplets charged under two gauge groups resembles the reduction of the $2d$ $\CN=(0,2)$ SQCD theory on the sphere in the presence of gauge fluxes, which we briefly discussed in section \ref{scompsp}.
It would be interesting to understand and derive orbifold reduction towards brane hyper-brick models corresponding to toric Calabi-Yau fivefolds in terms of sphere reductions, as we have done for SCQD-type theories in this paper.

\section{Discussion}

We introduced a convenient way to construct supersymmetric matrix models, also known as $0d$ $\CN=1$ ``quantum field theories,'' by compactification of $2d$ $\CN=(0,2)$ supersymmetric theories on a sphere with the half-topological twist. In this setup, one can  naturally derive quadrality relations among $0d$ theories from $2d$ Gadde-Gukov-Putrov triality.
Together with the results of \cite{Honda:2015yha, Gadde:2015wta}, this leads to an interesting unification of Seiberg-like dualities in even dimensions by twisted compactification. 
 Assuming that the topological twisting commutes with the RG flow in two dimensions, our results provides a field theory ``derivation'' of matrix-model quadrality, which can be understood as a property of a subsector (the zero-instanton sector) of $2d$ SQCD. In contrast, previous evidence for quadrality mostly relied on its string theory embedding \cite{Franco:2016qxh, Franco:2016tcm}.

It would be interesting to study such twisted compactifications more thoroughly, in a number of dimensions and with various amounts of supersymmetry. In the present context, it would be important to study the non-zero gauge flux sectors systematically. It would also be crucial to actually compute general observables in $0d$ SQCD. This is essentially a toy-model for the direct computation of the $2d$ $\CN=(0,2)$ half-BPS observables on the sphere, which encode rather important chiral algebras---see {\it e.g.} \cite{Gadde:2013lxa, Gadde:2014ppa, Dedushenko:2015opz, Dedushenko:2017osi}. We hope for this note to be a modest step toward that ambitious goal.

\acknowledgments{
We would like to thank Sangmin Lee, Brian Willett, Shing-Tung Yau and Youngbin Yun for enjoyable and helpful discussions. The work of D.G. is supported by POSCO TJ Park Science Foundation and by Samsung Science and Technology Foundation under Project Number SSTF-BA1402-08. D.G also gratefully acknowledge Korean Physical Society (KPS) and CERN for hospitality and support through CERN-Korea theory collaboration while part of this work was being carried out. }
\\

\appendix
\section{2d $\CN=(0,2)$ supersymmetry and the half-twist} \label{convention}

In this appendix, we review our notation for $2d$ $\CN=(0,2)$ supersymmetric gauge theories  \cite{Witten:1993yc}. For a thorough discussion of the half-topological twist, we refer to \cite{Closset:2015ohf} and references therein. There are three kinds of supermultiplets in $2d$ $\CN=(0,2)$ gauge theories. All component fields are assumed to be complex-valued unless specified otherwise. The multiplets are:

\paragraph{$\CN=(0,2)$ chiral multiplet.}
Its physical (on-shell) component fields are a boson $\phi$ and a right-moving Fermion $\psi_+$. 
\begin{align} \label{2d02chiral}
\Phi = (\phi\,, \psi_+ ) \,, \, \quad \qquad  \overline{\mathcal{D}}_+ \, \Phi =0 \,,
\end{align}
where $\overline{\mathcal{D}}_+$ is a supercovariant derivative.

\paragraph{$\CN=(0,2)$ Fermi multiplet.}
The only physical field is the left-moving fermion $\lambda_-$, which is a supersymmetry singlet in the free theory limit, and its charged conjugate. The off-shell multiplet also contains an auxiliary field $G$. A deformation of the chirality condition allows for a coupling to a holomorphic function $E(\Phi)$ of chiral superfields:
\begin{align}\label{2d02fermi}
\Lambda = (\lambda_-\,, G) \,, \, \quad \qquad  \overline{\mathcal{D}}_+ \, \Lambda =E(\Phi) \,.
\end{align}

\paragraph{$\CN=(0,2)$ Vector multiplet.}
It contains the real gauge boson $v^\mu$, complex gaugini $\chi_{-}$ and a real auxiliary field $D$ while the on-shell degree of freedoms are gaugini:
 \begin{align}\label{2d02vector}
V=(v_{\mu}, \chi_-,\b \chi_-,  D) \,.
\end{align}
 They couple to matter fields minimally through a supersymmetric completion of the gauge-covariant derivative. 
 
For each Fermi multiplet $\Lambda_I$, in addition to the holomorphic $E_I$-term mentioned above, it is possible to introduce another holomorphic term called $J^I(\Phi)$. 
The $\CN=(0,2)$ supersymmetry requires that $J$- and $E$-terms satisfy an overall constraint:
\begin{align}
\sum_I \Tr \left( E_I(\Phi) J^I(\Phi) \right)= 0 \,.
\end{align}

Integrating out the auxiliary fields $D_\alpha$, we obtain a familiar looking $D$-term potential (and its fermionic partners). For abelian theories, the potential takes the form
\begin{align}
V_D = \sum_\alpha \left( \sum_i Q_{\alpha}^i |\phi_i|^2 - t_\alpha \right)^2 \,,
\label{V_D}
\end{align}
where $t_\alpha$ are complexified Fayet-Iliopoulos (FI) parameters.
Integrating out the auxiliary fields $G_I$, we obtain what may be called an $F$-term potential, 
\begin{align}
V_F = \sum_I  \Tr\Big(|E_I(\phi)|^2 +  |J^I(\phi)|^2 \Big)\,,
\label{V_JE}
\end{align}
as well as Yukawa-like interactions between scalars and pairs of fermions.

\subsection*{Zero-modes on the twisted sphere}
For each dynamical degree of freedom in $2d$, one can analyze its spectrum and count how many zero-mode survives after sphere compactification, as follows. The spectrum of scalars $\phi_i$ in $\CN=(0,2)$ chiral multiplet $\Phi_i$ on sphere $S^2$ with magnetic flux $\n$ is determined by the eigenvalue problem:
\be\label{phi-zero-mode-eq}
-4 D_z D_{\b z} \phi = l_{\n_{\rho_i}} \phi \,, \quad \quad l_{\n_{\rho_i}} \geq 0 \,,
\ee
where $z, \b z$ denote the complexified frame indices.
The eigenfunctions are known as monopole spherical harmonics \cite{WU1976365}.
On the round $S^2$, the spectrum takes the form:
\begin{align}
l^{j}_{(i,k)} = j(j+1) - {{(\n_{\rho_i}-1)(\n_{\rho_i}+1)}\ov{4}} \,, \quad \quad j=j_0 \,, \, j_0+1 \,, \, \cdots \,,
\end{align}
with 
\be
j_0 (\n_{\rho_i}) = {|\n_{\rho_i}|-1\ov{2}} \,, \quad \quad \n_{\rho_i} = \rho_i(\m) - r_i +1~.
\ee
It is known that each eigenvalue has multiplicity $2j+1$. The zero-mode appears at $j=j_0$ with multiplicity $2j_0+1$. Those zero-modes, in fact, exist for any metric on the sphere with flux. They are solutions to the equation:
\be
D_{\b z}\phi=0~,
\ee
where $z$ is a complex coordinate on the sphere.
These scalar zero-modes are holomorphic sections of a line bundle $\CO(\n_{\rho_i}-1)$ over $\mathbb{P}^1$, and it is well known that there are $\n_{\rho_i}$ such modes if and only if $\n_{\rho_i} > 0$.

For the chiral fermions in $\CN=(0,2)$ multiplets, the spectrum is not well-defined by itself (as for any chiral fermion), but the number of zero-modes is again completely determined by the topology of the line bundles.
For right-moving fermions $\psi_{+i}$ in $\CN=(0,2)$ chiral multiplets $\Phi_i$, we have:
\begin{align}\label{psi-zero-mode-eq}
\text{(zero-mode equation for $\psi_{+i}$)} \rightarrow \begin{cases}
D_{\b z} \overline \psi_{+i} =0 \,, &  \;\; {\rm if} \;\; \n_{\rho_i} \geq 0 ~, \cr 
D_z \psi_{+i} = 0 \,, & \; \; {\rm if} \;\; \n_{\rho_i} < 0 ~.  \end{cases}
\end{align}
This leads to \eqref{zero modes count chiral}.
Similarly, for the left-moving fermions $\lambda_-, \b\lambda_-$ in fermi multiplets with $\n_{\rho_I}=\rho_I(\m) - r_I$, we have:
\begin{align}\label{lambda-zero-mode-eq}
\text{(zero-mode equation for $\lambda_{-I}$)} \rightarrow \begin{cases}
D_{\b z} \lambda_{- I} = 0 & \;\; {\rm if}  \;\; \n_{\rho_I} \geq 0 ~, \cr
D_{z} \overline{\lambda}_{-I} =0 & \; \;  {\rm if} \;\; \n_{\rho_I} < 0 ~,  \end{cases}
\end{align}
which gives us \eqref{zero modes count Fermi}. This zero-mode counting was previously discussed in  \cite{Closset:2015ohf}.


\bibliographystyle{JHEP}
\bibliography{mybib}


\end{document}